\documentclass[prd,aps,nofootinbib,11pt]{revtex4}
\usepackage{amsmath,graphicx,color,epsfig}

\begin{document}

\title{Partners  of $Z(4430)$ and Productions in $B$ Decays}

\author{Ying Li$^{a}$, Cai-Dian L\"u$^{b}$ and Wei Wang$^{b,c}$}
\affiliation{\it \small $^a$  Physics Department, Yantai University,
Yantai, 264005, P.R. China\\
\it \small $^b$  Institute of High Energy Physics, P.O. Box 918(4)
Beijing, 100049, P.R. China\\
\it \small $^c$ Graduate University of Chinese Academy of Sciences,
Beijing,
 100049, P.R. China }

\begin{abstract}
Recently, Belle Collaboration has reported a resonant state produced
in $B\to K \pi\psi'$, which is called $Z(4430)$. This state is
charged, so it can not be interpreted as an ordinary charmonium
state. In this paper, we   analyze the octet to which this particle
belongs and predict the masses of mesons in this octet. Utilizing
flavor  SU(3) symmetry, we study production rates in several kinds
of $B$ decays. The $\bar B^0\to Z_s^-\pi^+\to K^-\psi'\pi^+$ and
$B^-\to \bar Z_s^0 \pi^-\to K_S\psi'\pi^-$ decay channels,  favored
by Cabibbo-Kobayashi-Maskawa matrix elements, can have branching
ratios of ${\cal O}(10^{-5})$. This large branching ratio could be
observed at the running $B$ factories  to detect $Z_s$ particles
containing a strange quark.  We also predict large branching ratios
of the $Z$ and $Z_c$ ($\bar cc\bar c D, D=u,d,s$) particle
production rates in non-leptonic $B_c$ decays and radiative $B$
decays. Measurements of these decays at the ongoing $B$ factories
and the forthcoming Large Hadron Collider-b experiments are helpful
to clarify the mysterious $Z$ particles.
\end{abstract}

\maketitle

\section{Introduction}

Recently, there are many exciting discoveries on new hadron states
especially in the hidden-charm sector. Among these discoveries, the
most intriguing one is the new relatively narrow peak named
$Z(4430)$ found by   Belle Collaboration  in the invariant mass
spectrum of $\pi\psi^{\prime}$ in the decay mode $B\to
K\pi\psi^{\prime}$\cite{2007wga}. There is a large branching
fraction for the following decay chain:
\begin{eqnarray}
 {\cal BR}(\bar B^0 \to K^-Z^+(4430)) \times  {\cal BR}(Z^+(4430) \to \pi^+\psi') = (4.1 \pm 1.0({\rm stat})
 \pm1.3({\rm syst})) \times 10^{-5}.
 \end{eqnarray}
Mass and width of this particle are measured as:
\begin{eqnarray}
 m_Z&=&(4433\pm 4\pm 1)  \mathrm{MeV};   \nonumber  \\
 \Gamma_Z&=&(44^{+17+30}_{-13-11}) \mathrm{MeV}.
\end{eqnarray}
The most prominent characteristic is that it is electric charged,
that's to say, this new particle can not be described as an ordinary
charmonium state or a charmonium-like state such as $\bar ccg$. On
the other hand, this particle can decay to $\pi^+\psi'$ with a large
rate through strong interactions, so it involves at least four
quarks $c\bar cu \bar d $, though there is  not any further detailed
information on its inner dynamics at present.

In order to elucidate this particle, many theoretical studies
\cite{Maiani:2007wz,Rosner:2007mu,Meng:2007fu,Bugg:2007vp,
Qiao:2007ce,Gershtein:2007vi,Lee:2007gs} have been put forward. This
meson could be viewed as a genuine tetraquark state with diquark
anti-diquark $[cu][\bar c \bar d]$ content which has a large rate to
$\pi\psi'$ \cite{Maiani:2007wz}. Moreover based on QCD-string, two
different four-quark descriptions are proposed in
Ref.~\cite{Gershtein:2007vi}: one can be reduced to the ordinary
diquark-diquark picture and the other one can not. Besides this kind
of explanation, it has also been identified as the resonance of
$D_1(D_1')D^*$ \cite{Rosner:2007mu,Meng:2007fu} as its mass is close
to the thresholds of $D^*(2010)D_1(2420)$ and $D^*(2010)D_1'(2430)$.
Within this picture, the authors in Ref. \cite{Meng:2007fu} explored
the production of $\pi\psi$ and $\pi\psi'$. Short distance
contribution to $Z\to\pi\psi(\psi')$ is neglected and the main
contribution is from long distance re-scattering effect  via $D^*
D_1$. With proper parameters, they can successfully explain the much
larger production rate of $\pi\psi'$ than that of $\pi\psi$.
In Ref. \cite{Bugg:2007vp}, Bugg took this meson as a $D^*(2010)\bar
D (2420)$ threshold cusp. Recently, Qiao also tried to explain this
meson with the baryonium picture \cite{Qiao:2007ce}. Using the
technique of QCD sum rules, Lee $et. al$ calculated the masses of
this particle and its strange partner $Z_s$ in Ref.
\cite{Lee:2007gs}.

Whether or not these scenarios describe the true dynamics of
$Z(4430)$, this strange meson indeed plays an important role in the
charmonium spectroscopy. In the present paper, we do not intend to
give an explanation of this meson's  structure, but we want to
analyze its partners within SU(3) symmetry: the octet to which the
meson $Z(4430)$ belongs and the corresponding singlet meson. Up to
now, these is no experimental information on these mesons except
$Z^\pm$. The decays of $B$ meson provide a firm potential in
searching for these exotic mesons \cite{Rosner:2003ia,Bigi:2005fr},
just like the observed decay channel $B\to KZ\to K\pi\psi'$. We will
investigate the possibilities to detect these $Z$ mesons in
$B_q(q=u,d,s,c)$ decays. In doing this, we will analyze  decay
amplitudes with the assumption of SU(3) flavor symmetry: to
construct effective Hamiltonian using flavor SU(3) meson matrixes.
The decay amplitudes can also be studied by using  Feynmann
diagrams. In the discussion of $Z$ production with the graphic
technique, we only consider short-distance contributions and neglect
soft final state interactions. Specifically, the considered decays
are divided into three categories: Cabibbo-Kobayashi-Maskawa (CKM)
allowed non-leptonic decays; CKM suppressed non-leptonic decays;
radiative decays. The first kind of decays have similar branching
ratios with the observed $\bar B^{0}\to K^-\pi^{+}\psi'$, while the
second type of decays is suppressed by about one order in magnitude
and we will show that the running $B$ factories could hardly detect
this kind of decays. Radiative $B$ decay is a natural filter to
exclude the 0-spin mesons and furthermore this kind of process may
go through with a sizable branching ratio.

In the next section, we will analyze the octet of $Z$ meson within
flavor SU(3) symmetry and try to estimate their masses. We will
construct the effective Hamiltonian using meson matrices and then
use them to study the production rates of $Z$ mesons in $B$ decays.
In Sec.\ref{zccc}, we will introduce $Z_c$ meson which consists of
three charm quarks, together with a brief discussion on its
production in $B_c$ decays. We will summarize this note in the last
section.

\section{The octet and the singlet}

\begin{figure}[thb]
\begin{center}
\includegraphics[scale=0.4]{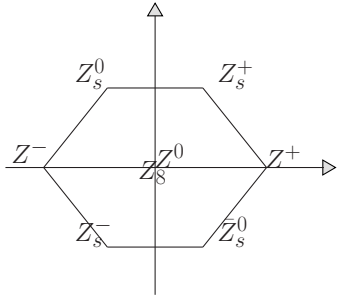}
\caption{Weight diagram for $Z$ meson octet.} \label{diagram:octet}
\end{center}
\end{figure}

Just as stated above, $Z(4430)$ involves at least four quarks in
constituent quark model, and there is an octet which $Z(4430)$
belongs to in flavor SU(3) symmetry. Generally, we can deduce the
particles in this octet using group theory: these particles, under
the name $Z^\pm$, $Z^0$, $Z_s^\pm$, $Z_s^0$, $\overline Z_s^0$ and
$Z_8$, are shown in Fig. \ref{diagram:octet}. Besides, there exists
one singlet meson called $Z_1$. In constituent quark model, quark
contents of these mesons are listed by:
\begin{align}
  Z^+ = c\bar c u  \bar d  ;~~~~
  Z^0 = \frac{1}{\sqrt{2}}c\bar c&( u  \bar u-d  \bar d)  ;~~~~
  Z^- =c\bar cd\bar u    ; \nonumber\\
  Z_s^+ = c\bar cu  \bar s ;~~~~
  Z_s^- =c\bar c  s\bar u  ;&~~~~
  Z_s^0 = c\bar cd  \bar s ;~~~~
  \overline Z_s^0 = c\bar cs\bar d ; \nonumber \\
  Z_8= \frac{1}{\sqrt{6}} c\bar c (u\bar u+d\bar d-2s\bar s);
  &~~~~Z_1 = \frac{1}{\sqrt{3}} c\bar c (u\bar u+d\bar d+s\bar s).
\end{align}
In reality, $s$ quark is slightly heavier than $u,d$ quark which is
one of the origins for SU(3) symmetry breaking. Accordingly, the
singlet $Z_1$ can mix with  eighth component of the octet $Z_8$, in
analogy with $\eta$ and $\eta'$. Physical particles, named
$Z_{\alpha}$ and $Z_{\beta}$, are mixtures of them and can be
expressed as:
\begin{eqnarray}
\left(%
\begin{array}{c}
  Z_\alpha \\
  Z_\beta \\
\end{array}%
\right)=\left(%
\begin{array}{cc}
  \cos\theta & \sin\theta \\
  -\sin\theta  & \cos\theta \\
\end{array}%
\right)\left(%
\begin{array}{c}
  Z_8 \\
  Z_1 \\
\end{array}%
\right).
\end{eqnarray}
The mixing angle $\theta$ can be determined through measuring decays
of these two particles   in future. For simplicity, we will assume
the mixing is ideal, i.e. $\theta=54.7^\circ$. In this case, the
quark contents are:
\begin{eqnarray}
 Z_{\alpha}=\frac{1}{\sqrt2}c\bar c(u\bar u+d\bar d),\;\;\; Z_\beta=c\bar cs\bar s.
\end{eqnarray}
All together, one can use the following meson matrix to describe
these mesons:
\begin{eqnarray}
 Z=\left(%
\begin{array}{ccc}
  \frac{Z^0}{\sqrt 2}+\frac{Z_8}{\sqrt 6} & Z^+    &Z_s^+ \\
  Z^-  & -\frac{Z^0}{\sqrt 2}+\frac{Z_8}{\sqrt 6}  &Z_s^0  \\
  Z_s^-    &\bar  Z_s^0 &  -\sqrt {\frac{2}{ 3}}Z_8\\
\end{array}%
\right)+\frac{Z_1}{\sqrt 3}\bf {1}.
\end{eqnarray}

With the quark contents given in the above, we are ready to estimate
masses of these particles. Isospin analysis predicts the equal
masses for the four mesons with neither open nor hidden strangeness:
$Z^\pm, Z^0, Z_\alpha$. For the mesons with a strange quark, the
mass differences between the lighter $u,d$ quarks and the heavier
$s$ quark are required. One can compare masses of $D^{*}$ and
$D_s^*$ to get some information: the mass of $D_s^*$ is $100$ MeV
larger than that of $D^*$. In  heavy quark limit $m_c\to \infty$,
the light system will not be affected by  different heavy quark
systems, thus we can simply assume a similar difference for $Z$
mesons which predicts the mass of $Z_s$ around $4533$ MeV. Because
the mass of newly observed $Z$ meson is not far from the threshold
of $D^*(2010)D_1(2420)$, $Z^-$ meson is regarded as the resonance of
$D^*\overline D_1(2420)$ \cite{Rosner:2007mu}. Under this mechanism,
we could give more precise predictions on the masses for other $Z$
mesons using experimental results for the $D^*$ and $D_1$ mesons.
Our results are listed in Tab.~\ref{Tab:mass} and uncertainties in
this table are from that of masses of the charmed mesons. In the
heavy quark limit, mesons with the same light system can be related
to each other. But if the $Z$ particles are viewed as tetra-quark
states, the effective strange quark mass in $Z$ could be different
from that in the usual mesons as the light systems in the two kinds
of particles are different. If $Z$ mesons are described as
molecules, probably they would not belong to a full SU(3) nonet and
the predicted masses may not be suitable. Currently, there is no
better solution and we will use this assumption in the present
study. The recent QCD sum rule study predicts the mass by
\cite{Lee:2007gs}:
\begin{eqnarray}
 m_{Z_s} = (4.70\pm0.06) {\rm GeV},
\end{eqnarray}
which is above the $D_s^*D_1$ and $D^*D_{s1}$ threshold by about 160
MeV. More experimental studies are required to test this
description.

\begin{table}
\begin{center}
\caption{$Z$ meson and its mass} \label{Tab:mass} \vspace{5mm}
\begin{tabular}{|l| c| c|c|c|  }
 \hline
  ~~~Meson                                      & Constituent Meson         & Mass($\mathrm{MeV})$   &Decay Mode
 \\ \hline
  $Z^+,Z^-,Z^0,Z_\alpha$                        & $D^*(2010)\bar D_1(2420)$ & $4432.3\pm1.7$         &$\psi'\pi/\eta(\eta'), \eta_c(2S)\rho/\omega$
  \\\hline
  $Z_s^+,Z_s^-,Z_s^0,\bar Z_s^0$&$D^*_s(2112)\bar D_1(2420)/D^*(2010)\bar D_{s1}(2536)$ &$4534.3\pm1.9/4535.35\pm1.0$&$\psi' K, \eta_c(2S)K^*$
  \\\hline
  $Z_\beta$                                     &$D^*_s(2112)\bar D_{s1}(2536)$&$4647.35\pm1.2$       &$\psi'\eta(\eta'), \eta_c(2S)\phi$
  \\ \hline
\end{tabular}
\end{center}
\end{table}

Experimentalists have observed the $Z$ particle through the $B\to
ZK$ with $Z\to \pi\psi'$. Assuming S-wave decay for $Z$ meson, the
quantum numbers can be determined as
$J^{PC}=1^{+-}$~\cite{Maiani:2007wz}.  In order to detect the other
$Z$ mesons, experimentalists will choose the proper final states to
re-construct them, thus the predictions on $Z$'s strong decays are
required. Using the flavor SU(3) symmetry and $Z\to \pi\psi'$, we
also list the strong decays of other $Z$ mesons in Tab.
\ref{Tab:mass}.  With the assumption $J^{PC}=1^{+-}$, another kind
of possible decay modes is $Z\to \eta_c(2S)V$~\cite{Maiani:2007wz},
where $V$ denotes a light vector meson.

In order to explore the production in $B$ decays, one can construct
the effective Hamiltonian at hadron level using  meson matrices
\cite{Savage:1989ub}. In the following, to construct the related
effective Hamiltonian, we will assume the flavor SU(3) symmetry. In
$B_{u,d,s}$ decays, the initial state $B=(B^-,\bar B^0,\bar B^0_s )$
forms an SU(3) anti-triplet. The transition at quark level is either
$b\to c\bar cs$ or $b\to c\bar cd$\footnote{If the $\bar cc$ quark
pair is generated from the QCD vacuum rather than directly produced
by the four-quark operator, this kind of contribution is expected to
be suppressed by $\alpha_s(2m_c)$ since there is at least one hard
gluon required to produce the $\bar cc$ quark pair.}, which is
described by the effective electro-weak Hamiltonian:
 \begin{eqnarray}
 H &=& \frac{G_{F}}{\sqrt{2}}
     V_{cb} V_{cD}^{*} \big[
     C_{1} ({\bar{c}}_{\alpha}b_{\beta} )_{V-A}
               ({\bar{D}}_{\beta} c_{\alpha})_{V-A}
  +  C_{2} ({\bar{c}}_{\alpha}b_{\alpha})_{V-A}
               ({\bar{D}}_{\beta} c_{\beta} )_{V-A}\Big] + \mbox{H.c.} ,
 \label{eq:hamiltonian}
\end{eqnarray}
where $D=d,s$. $\alpha$ and $\beta$ are color indices. The
transition $b\to c\bar cs$ is CKM favored: $V_{cb}V_{cs}^*\sim 1$,
while the $b\to c\bar cd$ transition is suppressed by
$|V^*_{cd}/V^*_{cs}|=\lambda=0.23$. To construct the effective
Hamiltonian at hadron level, only the flavor structures needs to be
concerned. The effective electro-weak Hamiltonian given in
Eq.~(\ref{eq:hamiltonian}) can also be written as an SU(3) triplet:
$H^i$ (i=1 (u), 2 (d), 3(s)), where the only non-zero elements are
$H^3=1$ for CKM favored decays $b\to c\bar cs$, and $H^2=1$ for CKM
suppressed channels $b\to c\bar cd$. The final state mesons can be
described by two nonet matrices: $Z$ and $M$. The effective
Hamiltonian at hadron level could be constructed as:
\begin{eqnarray}
 {\cal H}
  = {\cal A}_3 B_i H^i Z^k_l M^l_k
  +{\cal B}_3 B_i H^j Z^i_l M^l_j +{\cal C}_3 B_i H^j Z^k_j M^i_k
  +{\cal D}_3 B_i H^j Z^i_j M^l_l + {\cal E}_3 B_i H^j Z^l_l
  M^i_j,\label{eq:EH}
\end{eqnarray}
where the upper index labels rows and the lower labels columns.

\begin{figure}[thb]
\begin{center}
\includegraphics[scale=0.5]{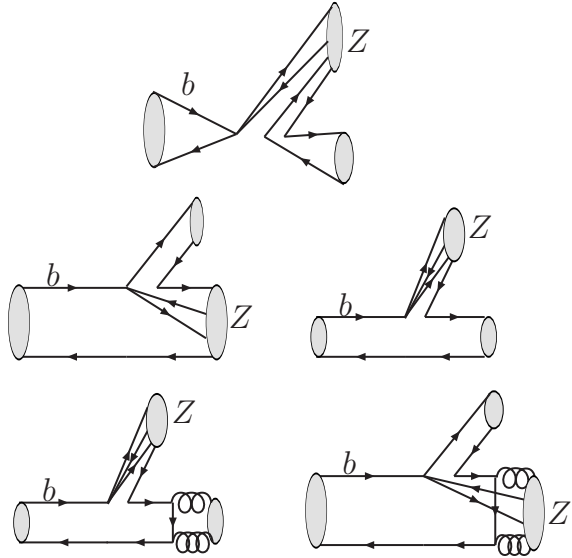}
\caption{Typical Feynman diagrams: annihilation (first row),
emission (second row) and gluonic diagrams (third row) of $Z$ meson
production in $B$ decays.} \label{diagram:Z-pro1}
\end{center}
\end{figure}

The above effective Hamiltonian can be related to Feynmann diagrams
with the one-to-one correspondence and the lowest order diagrams are
given in Fig.~\ref{diagram:Z-pro1}.  The second term in
eq.(\ref{eq:EH}) corresponds to the second diagram in
Fig.~\ref{diagram:Z-pro1} (called $Z$-recoiling diagram) in which
the spectator light quark in $B$ meson enters into the heavy $Z$
meson. If the spectator quark goes to the light meson, we call this
kind of diagram (the third one in Fig.~\ref{diagram:Z-pro1}) as the
$Z$-emission diagram which corresponds to the third term in the
effective Hamiltonian. In order to estimate relative sizes of these
terms, we have to analyze diagrams at quark level. Final state
mesons move very slowly and thus the gluon generating the $q\bar q$
quark pair is soft: $\alpha_s\sim {\cal O}(1)$. Thus after
integrating out high energy scales, decay amplitudes can be
expressed as matrix elements of a soft four-quark operator between
initial and final states. The first term in Eq.~(\ref{eq:EH})
corresponds to the annihilation diagram (the first one in
Fig.~\ref{diagram:Z-pro1}), as flavor indices of $B$ and $H$ in this
term are contracted with each other. This kind of diagram is
expected to be suppressed in two-body non-leptonic $B$ decays. But
here since the gluons are soft, decay amplitudes can also be
expressed as time-ordered products of a soft four-quark operator and
the ${\cal O}(1)$ interaction Hamiltonian which contains only soft
fields, thus this kind of contribution is comparable with
contributions from   the second and third terms in
Eq.~(\ref{eq:EH}). For SU(3) flavor singlet mesons $\eta_1$ and
$Z_1$, there are additional contributions which are given by the
last two terms in Eq.~(\ref{eq:EH}). One kind of typical Feynmann
diagram is also shown in Fig. \ref{diagram:Z-pro1} as the last two
diagrams and it is the contribution from the higher Fock states of
$\eta_1$ and $Z_1$. Even in charmless two-body $B\to
K(\pi)\eta(\eta')$ decays \cite{Williamson:2006hb}, this kind of
gluonic contribution is sizable. Here we do not have any implication
and thus one can not neglect it with any a priori.

\begin{table}
\begin{center}
\caption{SU(3) decomposition of $\Delta S=1$  $B_{u,d,s}$ decays,
whose decay amplitudes are proportional to
$V_{cb}V_{cs}^*$}\label{Tab:ampltiudesofbtos}
\begin{tabular}{|c| c|c|c|c|c|}
\hline
 Mode                     & ${\cal A}_3$   & ${\cal B}_3$ & ${\cal C}_3$ & ${\cal D}_3$  & ${\cal E}_3$  \\
\hline
 $B^-\to Z^0 K^-$         & $0$            & $1/\sqrt 2$  & $0$          & $0$           & $0$                  \\
 $B^-\to Z^- \bar K^0$    & $0$            & $1$          & $0$          & $0$           & $0$  \\
 $B^-\to Z_s^- \pi^0$     & $0$            & $0$          & $1/\sqrt 2$  & $0$           & $0$\\
 $B^-\to \bar Z_s^0 \pi^-$& $0$            & $0$          & $1$          & $0$           & $0$\\
 $B^-\to Z_8 K^-$         & $0$            & $1/\sqrt 6$  & $-\sqrt{2/3}$& $0$           & $0$ \\
 $B^-\to Z_1  K^-$        & $0$            & $1/\sqrt 3$  & $1/\sqrt 3$  & $0$           & $\sqrt 3$\\
 $B^-\to Z_s^- \eta_8$    & $0$            & $-\sqrt{2/3}$& $1/\sqrt 6$  & $0$           & $0$\\
 $B^-\to Z_s^- \eta_1$    & $0$            & $1/\sqrt 3$  & $1/\sqrt 3$  & $\sqrt 3$     & $0$ \\
  \hline
 $\bar B^0\to Z^+ K^-$             & $0$   & $1$          & $0$          & $0$           & $0$\\
 $\bar B^0\to Z^0 \bar K^0$        & $0$   & $-1/\sqrt 2$ & $0$          & $0$           & $0$\\
 $\bar B^0\to Z_s^- \pi^+$         & $0$   & $0$          & $1$          & $0$           & $0$\\
 $\bar B^0\to \bar Z_s^0 \pi^0$    & $0$   & $0$          & $-1/\sqrt 2$ & $0$           & $0$\\
 $\bar B^0\to Z_8 \bar K^0$        & $0$   & $1/\sqrt 6$  & $-\sqrt{2/3}$& $0$           & $0$ \\
 $\bar B^0\to Z_1  \bar K^0$       & $0$   & $1/\sqrt 3$  & $1/\sqrt 3$  & $0$           & $\sqrt3$ \\
 $\bar B^0\to \bar Z_s^0 \eta_8$   & $0$   & $-\sqrt{2/3}$& $1/\sqrt 6$  & $0$           & $0$\\
 $\bar B^0\to \bar Z_s^0 \eta_1$   & $0$   & $1/\sqrt 3$  & $1/\sqrt 3$  & $\sqrt 3$     & $0$\\
  \hline
 $\bar B^0_s\to Z^+_s K^-$         & $1$   & $1$          & $0$          & $0$           & $0$\\
 $\bar B^0_s\to Z^0_s \bar K^0$    & $1$   & $1$          & $0$          & $0$           & $0$\\
 $\bar B^0_s\to Z^-_s K^+$         & $1$   & $0$          & $1$          & $0$           & $0$\\
 $\bar B^0_s\to \bar Z^0_s K^0$    & $1$   & $0$          & $1$          & $0$           & $0$\\
 $\bar B^0_s\to Z^+ \pi^-$         & $1$   & $0$          & $0$          & $0$           & $0$\\
 $\bar B^0_s\to Z^- \pi^+$         & $1$   & $0$          & $0$          & $0$           & $0$ \\
 $\bar B^0_s\to Z^0 \pi^0$         & $1$   & $0$          & $0$          & $0$           & $0$\\%
 $\bar B^0_s\to Z_8 \eta_1$        & $0$   & $-\sqrt 2/3$ & $-\sqrt 2/3$ & $-\sqrt 2$    & $0$\\
 $\bar B^0_s\to Z_1 \eta_8$        & $0$   & $-\sqrt 2/3$ & $-\sqrt 2/3$ & $0$           & $-\sqrt 2$\\
 $\bar B^0_s\to Z_8 \eta_8$        & $1$   & $2/3$        & $2/3$        & $0$           & $0$\\
 $\bar B^0_s\to Z_1 \eta_1$        & $1$   & $1/3$        & $1/3$        & $1$           & $1$\\
  \hline
\hline
\end{tabular}
\end{center}
\end{table}

With the effective Hamiltonian given in Eq. (\ref{eq:EH}), we give
the decay amplitudes for the first kind of non-leptonic $B_{u,d,s}$
decay channels in Table~\ref{Tab:ampltiudesofbtos}. These decays are
induced by the CKM allowed transition $b\to c\bar cs$ and go through
with a large decay rate (typically the same order with the observed
$\bar B^{0}\to K^-\pi^{+}\psi^{\prime}$). The flavor SU(3) symmetry
implies the following relations for $b\to c\bar cs$ decays:
\begin{eqnarray}
 2{\cal BR}(B^-\to Z^0 K^-)&=& {\cal BR}(B^-\to Z^- \bar K^0)= {\cal BR}(\bar B^0\to Z^+ K^-)= 2{\cal BR}(\bar B^0\to Z^0\bar K^0),\\
 2{\cal BR}(B^-\to Z_s^-\pi^0)&=& {\cal BR}(B^-\to \bar Z_s^0 \pi^-)= {\cal BR}(\bar B^0\to Z_s^-\pi^+)= 2{\cal BR}(\bar B^0\to \bar
 Z_s^0\pi^0),\\
 {\cal BR}(B^-\to Z_\alpha K^-)&= & {\cal BR}(\bar B^0\to Z_\alpha\bar
 K^0),\;\;\;
 {\cal BR}(B^-\to Z_\beta K^-)=  {\cal BR}(\bar B^0\to Z_\beta \bar
 K^0),\\
 {\cal BR}(B^-\to Z_s^- \eta)&= & {\cal BR}(\bar B^0\to\bar Z_s^0
 \eta),\\
 {\cal BR}(\bar B_s^0\to Z_s^+ K^-)&=& {\cal BR}(\bar B_s^0\to Z_s^0\bar
 K^0),\;\;\;
 {\cal BR}(\bar B_s^0\to Z_s^- K^+)= {\cal BR}(\bar B_s^0\to \bar Z_s^0
 K^0),\;\;\;\\
 {\cal BR}(\bar B_s^0\to Z^+\pi^-)&=& {\cal BR}(\bar B_s^0\to
 Z^-\pi^+)= {\cal BR}(\bar B_s^0\to Z^0\pi^0),
\end{eqnarray}
where  mass differences and lifetime differences of $B$ mesons are
neglected which can not produce large corrections. Although all of
these decays are expected to go through with large branching
fractions, decay rates may differ from each other for distinct
coefficients. Two of the decays in the first line have been observed
experimentally, while the possibility to observe the other two
channels is a little smaller as the daughter meson $\pi^0$ from
$Z^0$ is relatively more difficult to measure. The decays in the
second line is contributed from the third term of effective
Hamiltonian given in Eq. (\ref{eq:EH}), which should also have
similar production rates. Among these four channels, $\bar B^0\to
Z_s^-\pi^+$ and $B^-\to \bar Z_s^0 \pi^-$ can have large branching
ratios and the final states ($K^-\psi'\pi^+$ or $K_S\psi'\pi^+$) are
easily to be measured on the experimental side. Thus measurements of
the $K\psi'$ invariant mass distribution in these two channels are
helpful to detect the $Z_s$ particles and determine  relative sizes
of ${\cal B}_3$ and ${\cal C}_3$. The other $B$ decays are less
possible to be measured in the running $B$ factories as either
$\pi^0$ or $\eta$ is produced in the final state. The forthcoming
LHC-b experiments and Super-B factories can measure these decays,
together with the $\bar B_s^0$ decays.

\begin{table}
\begin{center}
\caption{SU(3) decomposition of  $\Delta S=0$  $B_{u,d,s}$ decays,
whose decay amplitudes are proportional to $V_{cb}V_{cd}^*$. }
\label{Tab:ampltiudesofbtod}
\begin{tabular}{|c| c|c|c|c|c|c|  }
\hline
 Mode                     & ${\cal A}_3$   & ${\cal B}_3$ & ${\cal C}_3$ & ${\cal D}_3$  & ${\cal E}_3$  \\
\hline
 $B^-\to Z^0 \pi^-$       & $0$            & $1/\sqrt2$   & $-1/\sqrt 2$ & $0$           & $0$           \\
 $B^-\to Z^- \pi^0$       & $0$            & $-1/\sqrt2$  & $1/\sqrt2$   & $0$           & $0$           \\
 $B^-\to Z_s^- K^0$       & $0$            & $1$          & $0$          & $0$           & $0$           \\
 $B^-\to Z_s^0 K^-$       & $0$            & $0$          & $1$          & $0$           & $0$           \\
 $B^-\to Z_8  \pi^-$      & $0$            & $1/\sqrt6$   & $1/\sqrt6$   & $0$           & $0$           \\
 $B^-\to Z_1  \pi^-$      & $0$            & $1/\sqrt3$   & $1/\sqrt3$   & $0$           & $\sqrt3$      \\
 $B^-\to Z^- \eta_8$      & $0$            & $1/\sqrt6$   & $1/\sqrt6$   & $0$           & $0$           \\
 $B^-\to Z^- \eta_1$      & $0$            & $1/\sqrt3$   & $1/\sqrt3$   & $\sqrt3$      & $0$           \\
  \hline
 $\bar B^0\to Z^+ \pi^-$             & $1$ & $1$          & $0$          & $0$           & $0$\\
 $\bar B^0\to Z^- \pi^+$             & $1$ & $0$          & $1$          & $0$           & $0$\\
 $\bar B^0\to Z^0 \pi^0$             & $1$ & $1/2$        & $1/2$        & $0$           & $0$\\
 $\bar B^0\to Z_s^+ K^-$             & $1$ & $0$          & $0$          & $0$           & $0$\\
 $\bar B^0\to Z_s^0 \bar K^0$        & $1$ & $0$          & $1$          & $0$           & $0$\\
 $\bar B^0\to Z_s^- K^+$             & $1$ & $0$          & $0$          & $0$           & $0$\\
 $\bar B^0\to \bar Z_s^0 K^0$        & $1$ & $1$          & $0$          & $0$           & $0$\\
 $\bar B^0\to Z_8  \pi^0$            & $0$ & $-1/2\sqrt3$ & $-1/2\sqrt3$ & $0$           & $0$ \\
 $\bar B^0\to Z_1  \pi^0$            & $0$ & $-1/\sqrt6$  & $-1/\sqrt6$  & $0$           & $-\sqrt{3/2}$ \\
 $\bar B^0\to Z^0  \eta_8$           & $0$ & $-1/2\sqrt3$ & $-1/2\sqrt3$ & $0$           & $0$ \\
 $\bar B^0\to Z^0  \eta_1$           & $0$ & $-1/\sqrt6$  & $-1/\sqrt6$  & $-\sqrt{3/2}$ & $0$ \\
 $\bar B^0\to  Z_8 \eta_8$           & $1$ & $1/6$        & $1/6$        & $0$           & $0$\\
 $\bar B^0\to  Z_8 \eta_1$           & $0$ & $1/3\sqrt2$  & $1/3\sqrt2$  & $1/\sqrt2$    & $0$\\
 $\bar B^0\to  Z_1 \eta_8$           & $0$ & $1/3\sqrt2$  & $1/3\sqrt2$  & $0$           & $1/\sqrt2$\\
 $\bar B^0\to  Z_1 \eta_1$           & $1$ & $1/3$        & $1/3$        & $1$           & $1$\\
  \hline
 $\bar B^0_s\to Z^+_s \pi^-$       & $0$   & $1$          & $0$          & $0$           & $0$\\
 $\bar B^0_s\to Z^0_s \pi^0$       & $0$   & $-1/\sqrt2$  & $0$          & $0$           & $0$\\
 $\bar B^0_s\to Z^- K^+$           & $0$   & $0$          & $1$          & $0$           & $0$\\
 $\bar B^0_s\to Z^0 K^0$           & $0$   & $0$          & $-1/\sqrt2$  & $0$           & $0$\\
 $\bar B^0_s\to Z_8 K^0$           & $0$   & $-\sqrt{2/3}$& $1/\sqrt6$   & $0$           & $0$\\
 $\bar B^0_s\to Z_1 K^0$           & $0$   & $1/\sqrt3$   & $1/\sqrt3$   & $0$           & $\sqrt3$\\
 $\bar B^0_s\to Z_s^0 \eta_8$      & $0$   & $1/\sqrt{6}$ & $-\sqrt{2/3}$& $0$           & $0$\\
 $\bar B^0_s\to Z_s^0 \eta_1$      & $0$   & $1/\sqrt3$   & $1/\sqrt3$   & $\sqrt3$      & $0$\\
  \hline
\hline
\end{tabular}
\end{center}
\end{table}

For $\bar B^{0}\to K^- Z^{+} $, the heavy $b$ quark decays into
$c\bar c s$, and $q\bar q$ is produced from vacuum. Subsequently, $c
\bar c$, $q$ and the spectator $\bar u$ can be transferred into $Z$,
and the quarks left form a kaon. The other $Z$ states can also be
produced by selecting a different quark pair $q\bar q$ or changing
the $s$ quark by $d$ quark. We give the decay amplitudes for
non-leptonic $B_{u,d,s}$ decay channels induced by $b\to c\bar cd$
transition in Table~\ref{Tab:ampltiudesofbtod}. These decays are
suppressed by CKM matrix elements $|V_{cd}/V_{cs}|^2$. $B^{-}\to
Z_s^-K^0$ is one example of this kind of decays and the product
branching ratio is:
\begin{eqnarray}
 & &\mathcal{BR}[B^{-}\to Z_s^-K^0]\times  \mathcal{BR}[Z_s^-\to K^{-}\psi^{\prime}]\nonumber\\
 &=&|\frac{V_{cd}}{V_{cs}}|^2 \times\mathcal{BR}[\bar B^{0}\to Z^+(4430)K^-]\times \mathcal{BR}[Z^+(4430)\to\pi^{+}\psi^{\prime}]\nonumber\\
 &=&(2.3\pm0.6\pm0.7)\times10^{-6},
\end{eqnarray}
where $|V_{cd}|=0.23$ and $|V_{cs}|=0.957$ \cite{Yao:2006px}. The
uncertainties are from the experimental results for
$\mathcal{BR}[\bar B^{0}\to Z^+(4430)K^-]\times
\mathcal{BR}[Z^+(4430)\to\pi^{+}\psi^{\prime}]$. In the above
calculation, mass differences and lifetime differences are neglected
again. From the branching ratio for this decay chain, we can see
that this kind of process receives strong suppression. Furthermore,
the detection of $K_S$ is more difficult than $K^-$, thus it could
hardly be measured at the present two $B$ factories. The relations
for the $b\to c\bar cd$ decay can  be derived similarly using the
effective Hamiltonian which are also useful in searching for the $Z$
mesons:
\begin{eqnarray}
 {\cal BR}(B^-\to Z_s^- K^0)&=& {\cal BR}(\bar B_s^0\to Z_s^+ \pi^-)= 2{\cal BR}(\bar B_s^0\to Z_s^0 \pi^0),\\
 {\cal BR}(B^-\to Z_s^0 K^-)&=& {\cal BR}(\bar B_s^0\to Z^- K^+)= 2{\cal BR}(\bar B_s^0\to Z^0 K^0),\\
 {\cal BR}(B^-\to Z^-\pi^0 )&= & {\cal BR}( B^-\to Z^0\pi^-),\\
 {\cal BR}(\bar B^0\to Z^+ \pi^-)&=&  {\cal BR}(\bar B^0\to \bar Z_s^0 K^0),\;\;\;
  {\cal BR}(\bar B^0\to Z^- \pi^+)=  {\cal BR}(\bar B^0\to  Z_s^0\bar K^0),\\
 {\cal BR}(B^-\to Z_{\alpha,\beta} \pi^-)&= & 2{\cal BR}(\bar B^0\to\bar Z_{\alpha,\beta}\pi^0),\;\;\;
 {\cal BR}(B^-\to Z^- \eta(\eta'))=  2{\cal BR}(\bar B^0\to
 Z^0\eta(\eta')).
\end{eqnarray}

\begin{figure}[thb]
\begin{center}
\includegraphics[scale=0.5]{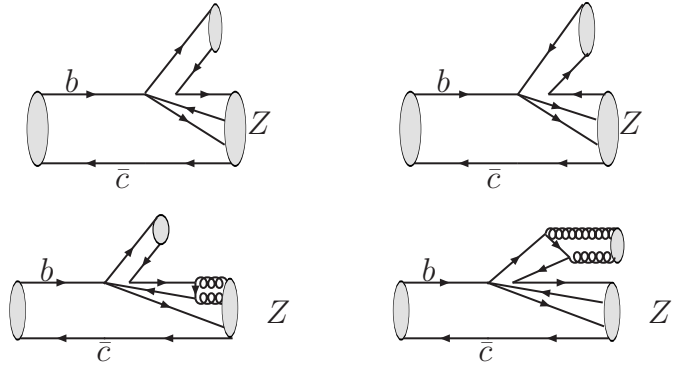}
\caption{Leading order Feynman diagrams of $Z$ meson production in
$B_c$ decays. }
 \label{diagram:ZproductioninBc}
\end{center}
\end{figure}

As pointed out in ref.\cite{xyz}, the study of charmonium like
states production in $B_c$ decays is easier. Here we also consider
the Z(4430) particle production in $B_c$ decays. In this case, the
spectator is a $\bar c$ quark, thus the initial state is very
simple: a singlet of flavor SU(3) group. But the effective
electro-weak Hamiltonian can form an octet: $3\otimes \bar 3=8
\oplus 1$. The effective Hamiltonian at hadron level can be written
by:
\begin{eqnarray}
 {\cal H}
  ={\cal A}_8  B H^i_j Z^j_l M^l_i + {\cal B}_8  B H^i_j Z^k_i M^j_k
  +{\cal C}_8 B H^i_j M^j_iZ^k_k + {\cal D}_8   B H^i_j Z^j_i M^l_l,
\end{eqnarray}
where the non-zero elements of the transition Hamiltonian are
$H^{2}_{1}=1$ for CKM allowed channels $b\to c\bar ud$ and
$H^{3}_{1}=1$ for CKM suppressed channels $b\to c\bar ud$ with a
factor $V_{us}^*/V_{ud}^*$. The corresponding Feynmann diagrams are
given in Fig.~\ref{diagram:ZproductioninBc}. The coefficients for
distinct contributions are given in Tab.~\ref{Tab:Bcdecays}. The CKM
matrix element for the decay channels induced by $b\to c\bar ud$ is
$V_{cb}V_{ud}^*$, which is in the same order with that of $B \to
KZ(4430)$: $V_{cb}V_{cs}^*$. Thus without any other suppressions,
these $B_c$ decays also have similar branching ratios (${\cal
O}(10^{-5})$) with $\bar B^0\to K^-Z^+\to K^-\pi^+\psi'$. The decays
in the second part of Tab.~\ref{Tab:Bcdecays} are suppressed by
$(V^*_{us}/V^*_{ud})^2$, which are expected to have smaller decay
rates (${\cal O}(10^{-6})$). Furthermore, the SU(3) symmetry implies
the following relations:
\begin{eqnarray}
 {\cal BR}(\bar B_c^-\to Z^0\pi^-)&=& {\cal BR}(\bar B_c^-\to
 Z^-\pi^0),\\
 2{\cal BR}(\bar B_c^-\to Z^0 K^-)&=& {\cal BR}(\bar B_c^-\to
 Z^- \bar K^0),\\
 2{\cal BR}(\bar B_c^-\to Z_s^- \pi^0)&=& {\cal BR}(\bar B_c^-\to
 \bar Z_s^0 \pi^-).
\end{eqnarray}

\begin{table}
\begin{center}
\caption{SU(3) decomposition of $B_{c}$ induced by $b\to c\bar ud$
(the first part) and $b\to c\bar us$ transitions (the second part) }
\label{Tab:Bcdecays} \vspace{5mm} {\small\begin{tabular}{|c|
c|c|c|c|c|c|  } \hline
    Mode                & ${\cal A}_3$  & ${\cal B}_3$ & ${\cal C}_3$ &${\cal D}_3$ \\
\hline
 $B_c^-\to Z_s^0 K^-$   & $0$           & $1$          & $0$          & $0$        \\
 $B_c^-\to Z_s^- K^0$   & $1$           & $0$          & $0$          & $0$        \\
 $B_c^-\to Z^0 \pi^- $  & $1/\sqrt2$   & $-1/\sqrt2$   & $0$          & $0$        \\
 $B_c^-\to Z^- \pi^0$   & $-1/\sqrt2$    & $1/\sqrt2$  & $0$          & $0$        \\
 $B_c^-\to Z_8 \pi^- $  & $1/\sqrt6$    & $1/\sqrt6$   & $0$          & $0$        \\
 $B_c^-\to Z_1 \pi^-$   & $1/\sqrt3$    & $1/\sqrt3$   & $\sqrt3$          & $0$   \\
 $B_c^-\to Z^-\eta_8$   & $1/\sqrt6$    & $1/\sqrt6$   & $0$          & $0$        \\
 $B_c^-\to Z^-\eta_1$   & $1/\sqrt3$    & $1/\sqrt3$   & $0$     & $\sqrt3$        \\
\hline
    Mode                & ${\cal A}_3$  & ${\cal B}_3$ & ${\cal C}_3$ &${\cal D}_3$ \\
\hline
 $B_c^-\to Z^0 K^-$         & $1/\sqrt2$       & $0$   & $0$          & $0$\\
 $B_c^-\to Z^- \bar K^0$    & $1$       & $0$          & $0$          & $0$\\
 $B_c^-\to\bar Z_s^0\pi^-$  & $0$       & $1$          & $0$          & $0$\\
 $B_c^-\to Z_s^- \pi^0$     & $0$& $1/\sqrt2$          & $0$          & $0$\\
 $B_c^-\to Z_8 K^- $      &$1/\sqrt6$& $-\sqrt{2/3}$   & $0$          & $0$\\
 $B_c^-\to Z_1 K^-$         & $1/\sqrt3$& $1/\sqrt3$   & $\sqrt3$          & $0$\\
 $B_c^-\to Z_s^-\eta_8$     & $-\sqrt{2/3}$& $1/\sqrt6$& $0$          & $0$\\
 $B_c^-\to Z_s^-\eta_1$     & $1/\sqrt3$& $1/\sqrt3$   & $0$     & $\sqrt3$\\
\hline
\end{tabular}}
\end{center}
\end{table}

\begin{figure}[thb]
\begin{center}
\includegraphics[scale=0.5]{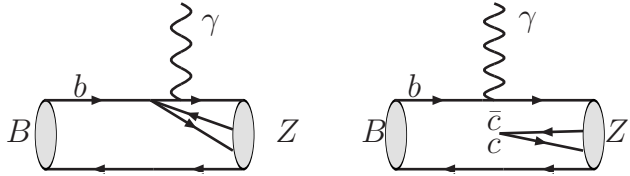}
\caption{$Z$ meson production via radiative decays. The right
diagram gives a smaller contribution as at least one hard gluon is
required.}
 \label{diagram:Z-pro4}
\end{center}
\end{figure}

Besides non-leptonic $B$ decays, $Z$ particles can also be produced
in radiative decays, as Fig.~\ref{diagram:Z-pro4} shows. Two-body
radiative decays can serve as a natural filter to exclude spin-0
candidates of Z particles, as the photon can only be transversely
polarized. But in order to predict the production rates, one has to
know $B\to Z$ transition form factors.  In the second diagram of
Fig.~\ref{diagram:Z-pro4}, in order to generate the $\bar cc$, we
require at least one hard gluon which will suppress the contribution
from this diagram.  Na\"ively thinking, the first diagram will also
be suppressed by the off-shell $c$ quark propagator, but since the
emitted photon is not  energetic (the total energy release is only
about $0.8$ GeV), the off-shellness is not large. This may imply the
following radiative $B$ decays could go through with considerable
rates:
\begin{eqnarray}
 B^-\to Z_s^-\gamma, \;\;\; \bar B^0\to \bar Z_s^0\gamma, \;\;\;\bar B_s^0\to
 Z_8(Z_1)\gamma,~~\bar B_c\to Z^-\gamma.
\end{eqnarray}

\section{$Z_{c}$ Particle}\label{zccc}

In the above, we have utilized the flavor SU(3) symmetry for light
quarks in $Z$ mesons. One can also replace one or two heavy $c$
quarks by heavier $b$ quarks which can predict $Z_b$ and $Z_{bb}$
mesons \cite{Cheung:2007wf}. Another attempt is to replace a light
quark by a heavy $c$ quark. With this replacement, we obtain three
$Z_c$ states which contain three heavy quarks and a light quark:
$\bar cc \bar c u$, $\bar cc \bar c d$ and $\bar cc \bar c s$,
together with their charge conjugates. We have to confess that this
replacement may change the internal dynamics, but here we assume the
same dynamics with $Z$. In this case, the masses can be obtained by
using mass differences of $c$ and $u,d,s$ quarks deriving from
masses of $\psi$ and $D^*(D^*_s)$. The rough predictions for masses
are around $5520$ and $5630$ MeV. If these mesons are viewed as the
resonance of $\psi$ and $D_1(D_{s1})$, their masses could be
predicted as $(5519.2\pm1.3)$ MeV, $(5519.2\pm1.3)$ MeV and
$(5632.3\pm0.6)$ MeV.

\begin{figure}
\begin{center}
\includegraphics[scale=0.4]{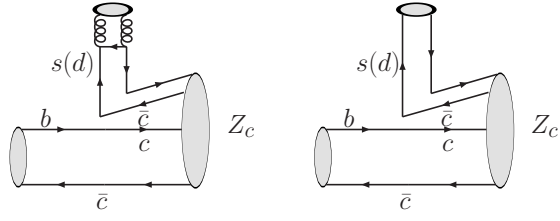}
\caption{Leading order Feynman diagrams of $Z_{c}$ meson production
in non-leptonic $B_c$ decays.} \label{diagram:Z-pro5}
\end{center}
\end{figure}

Because of the large mass of $Z_{c}$, these particles only appear in
$B_c$ meson decays. The corresponding effective Hamiltonian
responsible for non-leptonic decays is
\begin{eqnarray}
 {\cal H}
  ={\cal A}_3 BH^i Z_i M^k_k + {\cal B}_3 B H^i Z_j M^j_i,
\end{eqnarray}
where the first contribution ${\cal A}_3$ comes from the gluonic
diagram shown as the first one in Figure \ref{diagram:Z-pro5}; while
the second term  ${\cal B}_3$ comes from the $Z$-recoiling diagram
shown as the second one in Figure~\ref{diagram:Z-pro5}. These two
contributions give the following amplitudes for $8$ decays:
\begin{eqnarray}
 A(\bar B_c \to Z(\bar cc\bar c u) K^-) &=& A(\bar B_c \to Z(\bar cc\bar c d)\bar K^0)
 ={\cal B}_3,\label{28}\\
 A(\bar B_c \to Z(\bar cc\bar c s) \eta_8) &=&-\sqrt{\frac{2}{3}}{\cal B}_3\\
  A(\bar B_c \to Z(\bar cc\bar c s)\eta_1)&=&\sqrt3 {\cal A}_3+\frac{1}{\sqrt3}
 {\cal B}_3,\label{30}\\
 A(\bar B_c \to Z(\bar cc\bar c u) \pi^-) &=&-\sqrt2 A(\bar B_c \to Z(\bar cc\bar c d)
 \pi^0)=A(\bar B_c \to Z(\bar cc\bar c s)
 K^0)={\cal B}_3,\label{31}\\
 A(\bar B_c \to Z(\bar cc\bar c d) \eta_8) &=&\frac{1}{\sqrt6}{\cal B}_3\\
  A(\bar B_c \to Z(\bar cc\bar c d)\eta_1)&=&\sqrt3 {\cal A}_3+\frac{1}{\sqrt3}
  {\cal
 B}_3.\label{33}
\end{eqnarray}
The first 4 decay channels shown in eq.(\ref{28}-\ref{30}) are
induced by $b\to c\bar cs$ at quark level and thus have branching
ratios of order ${\cal O}(10^{-5})$, while the other decays shown in
eq.(\ref{31}-\ref{33}) are suppressed by $|V^*_{cd}/V^*_{cs}|^2$,
which have smaller decay rates (${\cal O}(10^{-6})$).

\section{Summary}

Belle Collaboration has reported a  resonance named $Z(4430)$, which
consists of at least four quarks in constituent quark model. In this
note, we analyze the octet to which this $Z$ meson belongs and the
corresponding singlet meson. Using the picture that the $Z$ mesons
are the resonances of $D^*$ and $D_1$ mesons, we estimate the masses
of these mesons. We investigate the production in non-leptonic
$B_{u,d,s}$ decays by constructing the effective Hamiltonian using
flavor SU(3) meson matrices. The transition at quark level is either
$b\to c\bar cs$ or $b\to c\bar cd$, where the former one is CKM
favored and the latter is suppressed by $|V^*_{cd}/V^*_{cs}|^2$.
Thus the considered non-leptonic decays have either similar
branching ratios with the observed decay $B^\pm\to Z^\pm \bar K^0$
($10^{-5}$) or smaller branching ratios ($10^{-6}$) as shown in the
text. Utilizing the SU(3) symmetry, we also obtain many relations
for various decay channels.  Measurements of the $K\psi'$ invariant
mass distribution in $\bar B^0\to Z_s^-\pi^+\to K^-\psi'\pi^+$ and
$B^-\to \bar Z_s^0 \pi^-\to K_S\psi'\pi^-$ are helpful to detect the
$Z_s$ particles and determine the relative size of ${\cal B}_3$ and
${\cal C}_3$. We also study the production rates in non-leptonic
$B_c$ decays and radiative $B$ decays in a similar way. Replacing a
light $u,d,s$ quark by a heavy $c$ quark, we get three states which
have three heavy quarks. The masses and the production in $B_c$
decays are also discussed. Measurements of all these decays at the
present $B$ factories and the forthcoming LHC-b experiments will
help us to clarify the new $Z$ particles.
\section*{Acknowledgements}

This work is partly supported by National Science Foundation of
China under Grant No.10735080 and 10625525 and by Foundation of
Yantai University under Grant No.WL07B19. We would like to
acknowledge F.-K. Guo, C-Z Yuan and S. Zhou for valuable
discussions.

\end{document}